\documentstyle[amssymb,multicol,epsfig,rotate,aps]{revtex}

\begin{document}
\draft

\title{Spin precession in disordered systems:\\
Anomalous relaxation due to heavy tailed field distributions} 

\author{Falk Scheffler$^{(1)}$ and Philipp Maass$^{(1,2)}$}

\address{$^{(1)}$ Fachbereich Physik, Universit\"at Konstanz, 78457
  Konstanz, Germany\\ $^{(2)}$ Institut f\"ur Physik, TU Ilmenau, 98684
  Ilmenau, Germany}

\date{October 26, 2001}

\maketitle

\begin{abstract}
  We investigate spin precession in the presence of randomly
  distributed field sources. Their fields ${\bf h}_i$ reorient by
  thermally activated transitions and decrease as $r^{-\mu}$ with the
  distance from the spin probe. Based on analytical calculations and
  scaling arguments we find that the polarization decay of a spin
  ensemble exhibits a rich behavior characterized by stretched
  exponenentials and power laws with exponents depending on $\mu$ and
  the dimension $d$. The anomalous relaxation laws result from heavy
  tailed local field distributions and are verified by computer
  simulations. Implications for experiments are pointed out.
\end{abstract}

\pacs{PACS numbers: 76.75.+i, 76.20.+q, 76.60.Es }

\begin{multicols}{2}
Many experimental probes rely on a precession of a spin ${\bf S}$ in
an external field ${\bf H}$,
\begin{equation}
\frac{d{\bf S}}{dt}=
{\bf S}\times{\bf H}\,.
\label{eq:precession}
\end{equation}
Examples are nuclear and electron magnetic resonance (NMR, ESR), muon
spin relaxation ($\mu$SR), $\beta$NMR, and quantum optical
measurements, where transitions in two--level systems can effectively
be described by an equation of type (\ref{eq:precession}). In
disordered systems the field $H$ generally exhibits both spatial and
temporal fluctuations and the relaxation of an initially polarized
spin ensemble is of interest. While traditionally this relaxation
dynamics is studied for Gaussian stochastic processes ${\bf H}(t)$
\cite{Kubo/Toyabe:1967,Uemura/etal:1985}, more complex stochastic
processes became of interest recently (see e.g.\ 
\cite{Keren/etal:1996}). Here we will focus on systems, where the
second moment $\langle H^2\rangle$ of the field distribution diverges.
These situations occur, when the field ${\bf H}=\sum_i {\bf h}_i$
results from randomly distributed sources $i$ in $d$ dimensions with a
spatial field dependence $h_i\sim r_i^{-\mu}$, $\mu>d/2$ (for dipolar
fields, in particular, $\mu=3$).

As an example of practical importance we focus on $\mu$SR in
disordered systems of single domain ferromagnetic particles
\cite{Bewley/Cywinski:1998,Jackson/etal:2000}. In these systems the
clusters perform thermally activated transitions between certain easy
magnetization directions with a rate $\nu$, which lead to fluctuations
of the magnetic field at the muon site. We will show in this Letter
that these fluctuations give rise to a rich anomalous relaxation
behavior due to the fact that the random spatial distribution of the
cluster moments leads to L\'evy type local field distributions.
Dependent on how the reorientation rate $\nu$ compares with the
characteristic width $W$ of the field distribution and dependent on
the number of possible orientations of the cluster moments, we find
very different relaxation scenarios. The long time relaxation is given
by either power laws or stretched exponentials, where the exponents
depend on both $\mu$ and $d$. The slow relaxations occur even in the
absence of cluster interaction effects and in this respect should be
contrasted to the relaxation found in spin glass systems
\cite{Keren/etal:1996} or related disordered systems
\cite{Lierop/Ryan:2001} with strongly interacting components.

To be specific, we consider the following model. We place a spin ${\bf
  S}$ at the origin of a $d$--dimensional system that contains
randomly oriented point-like clusters with number density $n$ at
random positions. A cluster with moment ${\bf m}$ and position ${\bf
  r}$ is assumed to induce a field contribution ${\bf h}={\bf
  m}/r^\mu$ at the probe site. Each moment ${\bf m}$ changes its
orientation to a set of possible other orientations with the rate
$\nu$. In particular we study two situations: In the first case only
the directions ${\bf m}$ and $-{\bf m}$ are possible (uniaxial case),
while in the second case there are four additional orientations
perpendicular to ${\bf m}$ corresponding to a cubic symmetry
(multiaxial case). Initially the spin is polarized in the
$z$--direction, ${\bf S}=(0,0,1)$. The task is to solve
eq.~(\ref{eq:precession}) for a given cluster configuration and a
certain realization of the cluster reorientation process and to
average this solution over all possible realizations. By finally
averaging over all cluster configurations we obtain the spin
polarization $\langle S_z(t)\rangle$ at time $t$ as measured in
experiment. In the following we will discuss the relaxation behavior
for the generic situation $\mu\!>\!d/2$ \cite{mu-d-comm}.

We start out by focusing on the time regime $t\ll\nu^{-1}$, where the
field ${\bf H}$ can be viewed to be static, and the solution of
eq.~(\ref{eq:precession}) reads
$S_z(t)=(H_z^2/H^2)+[1-(H_z^2/H^2)]\cos(Ht)$.  By an exact calculation
we obtain for the probability density $\psi({\bf H})$ of the local
field ${\bf H}$
\begin{equation}
\psi({\bf H})=\frac{1}{2\pi
  W^2H}\,{\rm Re}\,L'_{\frac{d}{\mu},0}\left(\frac{H}{W}\right)\,,
\label{eq:hdistribution}
\end{equation}
where $Re L'_{\alpha,0}(u)$ denotes the real part of the derivative of
the L\'evy stable law $L_{\alpha,0}(u)=(2\pi)^{-1}\int dk
\exp(-iku-|u|^\alpha)$ to the index $(\alpha,0)$ (see e.g.\ 
\cite{Bouchaud/Georges:1990}); the characteristic width
$W\!=\!C_Wmn^{\mu/d}$ is given by the field associated with the mean
distance $n^{-1/d}$ of the clusters times a constant \cite{c-comm}.
For large $H$, $4\pi H^2\psi({\bf H})\!\sim\!C_\psi
W^{-1}(H/W)^{-1-d/\mu}$, implying that $\langle H^2\rangle$ does not
exist.  Averaging $S_z(t)$ over $\psi({\bf H})$ eventually yields

\end{multicols}
\begin{figure}[t]
\epsfig{file=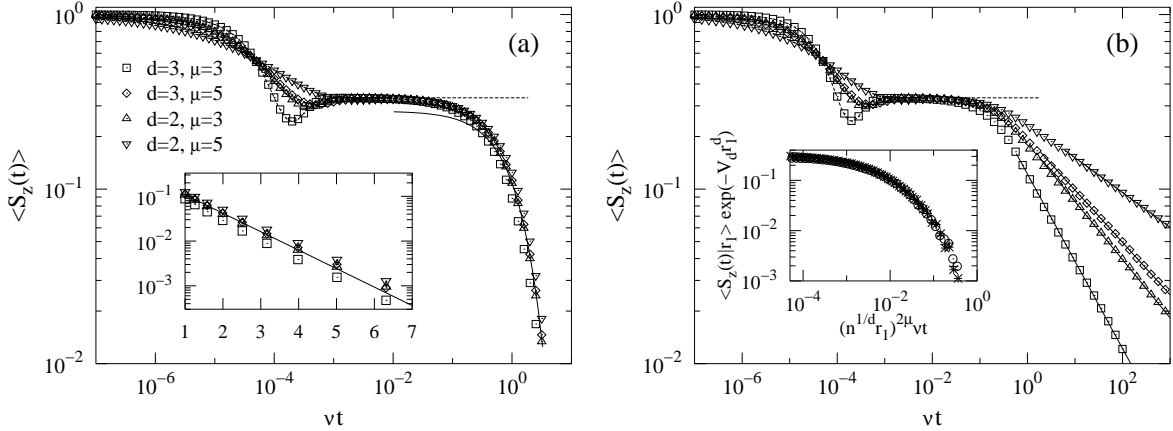,height=6cm}
\vspace*{0.1cm}
\caption{Spin polarization $\langle S_z(t)\rangle$ as a function of
  $\nu t$ in the slowly fluctuating case ($\nu/W\!=\!10^{-3}$) for
  {\it (a)} multiaxial and {\it (b)} uniaxial cluster moments, and
  several $\mu$ and $d$. The symbols refer to the simulations and
  their assignment is the same in both figures.  The dashed lines
  refer to the exact result (\ref{eq:sz(t)-static}), while the solid
  lines are fits according to the long-time behaviors
  (\ref{eq:sz(t)-slow-multiaxial},\ref{eq:sz(t)-slow}). The inset in
  {\it (a)} shows, on a semi-logarithmic scale, the exponential
  long-time relaxation of $\langle S_z(t)\rangle$ vs.\ $\nu t$ that is
  almost independent of $d$ and $\mu$ (the solid line is drawn as a
  guide for the eye). The inset in {\it (b)} demonstrates the scaling
  (\ref{eq:sz(t|r_1)-slow}) for 4 different radii $r_1\ll n^{-1/d}$,
  $r_1=1.0$ ($+$), 1.5 ($\times$), 2.0 ($*$), and 2.5 ($\circ$) in the
  case $\mu=d=3$, $n=0.01$.}
\label{fig:slow}
\end{figure}
\begin{multicols}{2}

\begin{equation}
\langle S_z(t)\rangle=
\frac{1}{3}+\frac{2}{3}\left[1\!-\!\frac{d}{\mu}(Wt)^{d/\mu}\right]
\exp\left[-(Wt)^{d/\mu}\right]\,.
\label{eq:sz(t)-static}
\end{equation}
For $d=\mu$, i.e.\ in particular for dipolar fields in $d=3$, one
recovers the Lorentzian Kubo--Toyabe function \cite{Kubo/Toyabe:1967}.
As shown in Fig.~\ref{fig:slow} for different $\mu$ and $d$, the
results from our simulations agree with eq.~(\ref{eq:sz(t)-static})
for $\nu t\ll1$. Laws of type (\ref{eq:sz(t)-static}) have been used
in the literature to describe anomalous $\mu$SR line-shapes with
$d/\mu\!\ne\!1,2$ that neither follow a Lorentzian ($d/\mu\!=\!1$) nor
Gaussian ($d/\mu\!=\!2$) behavior (see e.g.\ \cite{Wu/etal:1994}). We
note, however, that (\ref{eq:sz(t)-static}) is an exact result and
should not be confused with an effective ``power Kubo--Toyabe
function'' \cite{Crook/Cywinski:1997} that serves as a fitting
function.

In the dynamic regime $t\gg\nu^{-1}$ we distinguish between the two
cases of slowly or rapidly fluctuating cluster moments, where $\nu\ll
W$ or $\nu\gg W$, respectively. In both cases we employ scaling
arguments to derive the typical decay rates $\Gamma$ of the spin
polarisation. To tackle the problem of averaging over spatial cluster
configurations, we consider subensembles of configurations that are
specified by fixing the distances of the clusters closest to the spin
probe.  This concept is motivated by the hierarchy implied by the
L\'evy statistics, which for the field distribution
(\ref{eq:hdistribution}) means that the $n$th nearest cluster gives a
contribution of order $n^{\mu/d}$ times smaller than the closest
cluster (see e.g.\ \cite{Embrechts/etal:1997}).

Let us begin with the case $\nu\ll W$ of slowly fluctuating cluster
moments, where for the relevant cluster configurations the field ${\bf
H}$ has a magnitude $H\gg\nu$ (other configurations have an
exponentially small weight). In a time interval of order $\nu^{-1}$
then, the spin precesses many periods around the local field, whereby
$S_z(t)$ oscillates around a mean value $\bar S_z(t)$.  The changes of
$\bar S_z(t)$ averaged over many realizations of the cluster dynamics
determine the decay of spin polarization.

In the multiaxial case, significant changes of ${\bf H}$, which occur
in a time of order $\nu^{-1}$, alter the axis of precession and $\bar
S_z(t)$ relaxes with a rate proportional to $\nu$. Hence we expect a
simple exponential decay
\begin{equation}
\langle S_z(t)\rangle\sim\exp(-cst.\,\nu t)\,,
\label{eq:sz(t)-slow-multiaxial}
\end{equation}
which is confirmed by our simulations shown in Fig.~\ref{fig:slow}a.

The uniaxial case is more subtle. To see this, we decompose the field
${\bf H}$ into the contribution ${\bf h}_1={\bf m}/r_1^\mu$ from the
nearest cluster at distance $r_1$ and the contribution ${\bf H}_1$
from the other clusters, ${\bf H}={\bf h}_1+{\bf H}_1$. In the
subensemble of all cluster configurations with given $r_1$, the
variance of ${\bf H}_1$ is
\begin{equation}
\langle H_1^2|r_1\rangle=C_H\,h_1^2\,\left(\frac{h_1}{W}\right)^{-d/\mu}\,.
\label{eq:h1-var}
\end{equation}
For $r_1\!\gg\!n^{-1/d}$, $h_1/W\!\ll\!1$, and ${\bf H}_1$ dominates
over ${\bf h}_1$. Hence one encounters the same physical situation as
in the multiaxial case. For small $r_1\ll n^{-1/d}$, however, ${\bf
h}_1$ is dominant, so that changes ${\bf h}_1\to -{\bf h}_1$
essentially revert the direction of precession and leave $\bar S_z(t)$
unchanged.

In this situation of small $r_1\ll n^{-1/d}$ the presence of the
contribution ${\bf H}_1$ causes the axis of the field ${\bf H}$
(irrespective of its direction) to wobble around the $\pm {\bf
h}_1$-axis with the rate $\nu$ and an angular amplitude of order
$H_1/h_1$.  The wobbling motion together with the much faster
precession leads to a diffusive type of motion of $\bar S_z(t)$ with a
diffusion rate $\Gamma\sim(H_1/h_1)^2\nu$. 

To extract the asymptotic relaxation of the spin polarization we
consider the subensemble of all cluster configurations with fixed
distances $r_1$ and $r_2$ of the nearest and second nearest cluster to
the spin probe.  In the configurations of this subensemble we can
decompose ${\bf H}_1$ into ${\bf h}_2$ and ${\bf H}_2$, where
$h_2\!=\!m/r_2^\mu$ and $\langle H_2^2|r_2\rangle$ satisfies
(\ref{eq:h1-var}) with $h_1$ replaced by $h_2$. Accordingly, for
$r_1\!<\!r_2\!\lesssim\! n^{-1/d}$, $H_1^2\!\sim\! m^2/r_2^{2\mu}$ and
$\Gamma\!\equiv\!\Gamma(r_1,r_2)\!\propto\!(r_1/r_2)^{2\mu}\nu$, while
for $r_2\!\gtrsim\!n^{-1/d}$,
$H_1^2\!\sim\!W^{d/\mu}(m/r_2^\mu)^{2-d/\mu}$ and
$\Gamma(r_1,r_2)\!\propto\!r_1^{2\mu}W^{d/\mu}(m/r_2^\mu)^{2-d/\mu}\nu$.
Writing $\langle S_z(t)|r_1,r_2\rangle\sim\exp[-\Gamma(r_1,r_2)t]$ in
the subensemble with given $r_1$ and $r_2$, we can average over the
probability density
$\phi_2(r_2|r_1)\!=\!S_dnr_2^{d-1}\exp[-V_dn(r_2^d-r_1^d)]$ of $r_2$
($r_1\le r_2<\infty$) to obtain \cite{slow-comm}
\begin{equation}
\langle S_z(t)|r_1\rangle\sim
\exp\left\{V_dnr_1^d
\!-\!cst.\,\Bigl[(n^{1/d}r_1)^{2\mu}\nu t\Bigr]^{d/2\mu}\right\}
\label{eq:sz(t|r_1)-slow}
\end{equation}
for $\nu t\!\gg\!1$ (and $r_1\!\ll\! n^{-1/d}$).  We have verified this
prediction for various $\mu$ and $d$ by our simulations. One example
(for $\mu=d=3$) is shown in the inset of Fig.~\ref{fig:slow}b.

Final averaging over the probability density
$\phi_1(r_1)=S_dnr_1^{d-1}\exp[-V_dnr_1^d]$ of $r_1$ yields
\begin{equation}
\langle S_z(t)\rangle\sim (\nu t)^{-d/2\mu}\,.
\label{eq:sz(t)-slow}
\end{equation}
This slow power law decay is in marked contrast to the exponential
decay in the multiaxial case and it is verified in
Fig.~\ref{fig:slow}b by our simulations.

Next we discuss the case $\nu\gg W$ of rapidly fluctuating cluster
moments. The field ${\bf H}$ in the relevant cluster
configurations now has a magnitude $H\ll\nu$ and the spin rotates only by
a small angle in a time interval of order $\nu^{-1}$. This means that
the concept of a mean value $\bar S_z(t)$ is not useful any longer,
since the phase of the precession matters. Reorientations of ${\bf
  h}_1$ are effective for the spin relaxation both in the presence of
uniaxial and multiaxial cluster moments.

\begin{figure}[t!]
\hspace*{-0.7cm}
\epsfig{file=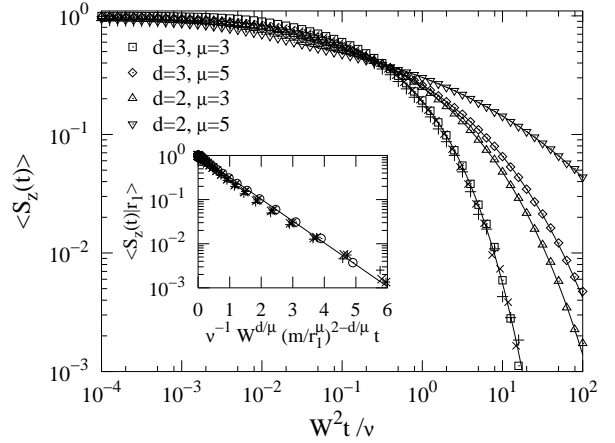,height=6cm}
\vspace*{0.2cm}
\caption{Spin polarization $\langle S_z(t)\rangle$ as a function of
  $W^2t/\nu$ in the case of rapidly fluctuating cluster moments
  [$\nu/W=10$ ($\square$), 50 ($+$), and 100
  ($\times$) for $d\!=\!\mu\!=\!3$, and $\nu/W=10$ for the three other
  combinations of $d$ and $\mu$]. Data points refer to the simulations
  and the solid lines are fits according to eq.~(\ref{eq:sz(t)-fast}).
  The inset shows the exponential decay of $\langle S_z(t)|r_1\rangle$
  and the scaling as discussed in the text for 4 different radii
  $r_1\gtrsim n^{-1/d}$, $r_1=6.5$ ($+$), 7.0 ($\times$), 7.5 ($*$),
  and 8.0 ($\circ$) in the case $\mu=d=3$, and $n=0.01$
  (the solid line is drawn as a guide for the eye).}
\label{fig:fast}
\end{figure}

The small angular changes of the spin lead again to a diffusive type
of motion of $S_z(t)$. In time $\nu^{-1}$ the angular change is of
order $H/\nu$ and the corresponding diffusion rate $\Gamma\sim
(H/\nu)^2\nu$. Decomposing the field ${\bf H}\!=\!{\bf h}_1+{\bf H}_1$
as before, and taking into account the dominant contributions we thus
find $\Gamma\equiv\Gamma(r_1)\!\propto\!\nu^{-1}m^2/r_1^{2\mu}$ for
$r_1\!\lesssim\!n^{-1/d}$ and $\Gamma(r_1)\!\propto\!\nu^{-1}W^{d/\mu}
(m/r_1^\mu)^{2-d/\mu}$ for $r_1\!\gtrsim\!n^{-1/d}$ [cf.\ 
eq.~(\ref{eq:h1-var})]. We then write $\langle
S_z(t)|r_1\rangle\!\sim\!\exp[-\Gamma(r_1)t]$ for $\nu t\!\gg\!1$ and
$r_1\!\gg\!(m/\nu)^{1/\mu}$ (for $r_1\!\ll\!(m/\nu)^{1/\mu}$,
$h_1\gg\nu$, i.e.\ one encounters a situation corresponding to the
case of slowly fluctuating cluster moments). This exponential decay of
$\langle S_z(t)|r_1\rangle$ is demonstrated in the inset of
Fig.~\ref{fig:fast} for $d\!=\!\mu\!=3$ in the regime
$r_1\!>\!n^{-1/d}$.  By averaging over $r_1$ we finally obtain
\begin{equation}
\langle S_z(t)\rangle\sim
\exp\left[-cst.\,\left(\nu^{-1}W^2t\right)^{d/2\mu}
\right]\,.
\label{eq:sz(t)-fast}
\end{equation}
To perform the average we have used a saddle point approximation,
where analogous comments apply as given in
\cite{slow-comm}. Figure~\ref{fig:fast} confirms both the scaling with
$(W^2t/\nu)$ and the stretched exponential decay for the same $\mu$
and $d$ values as in Fig.~\ref{fig:slow}. In the uniaxial case the
stretched exponential decay (\ref{eq:sz(t)-fast}) will, at long times,
be masked by the much slower power law decay (\ref{eq:sz(t)-slow})
that stems from the rare configurations with $h_1=m/r_1^\mu\gg\nu$.

In summary we have shown that spin precession in the presence of
randomly distributed and fluctuating field sources leads to an
anomalous relaxation of an initially polarized spin probe, which is
characterized by stretched exponentials
[eqs.~(\ref{eq:sz(t)-static},\ref{eq:sz(t)-fast})] or power laws
[eq.~(\ref{eq:sz(t)-slow})]. The deviation from a simple exponential
decay are caused by L\'evy type local field distributions
[eq.~(\ref{eq:hdistribution})].  These render a treatment in terms of
Gaussian processes impossible but allowed us to perform an analysis
based on subensembles of cluster configurations that are defined with
respect to the most dominant contributions to the local field, i.e.\
the field sources closest to the spin probe.

It is important to stress that a simple mean field type description of
the relaxation process would fail, as it was already pointed out by
Uemura {\it et al.} \cite{Uemura/etal:1985} in the case $\mu=d=3$.
In such a mean field description one might employ a ``strong collision
approximation'' \cite{kubo:sc}, where the field ${\bf H}$ at the probe
site is drawn anew from (\ref{eq:hdistribution}) with the rate $\nu$
(thereby neglecting the fluctuations in the spatial cluster
configurations). By scaling arguments similar to those outlined above
one can show that this approach leads, for $t\gg\nu^{-1}$, to an
exponential relaxation $\langle S_z(t)\rangle\sim\exp(-\Gamma_{\rm
  mf}t)$ both in the cases of slowly and rapidly fluctuating cluster
moments and irrespective of whether the clusters posses only one easy
axis or more. For $\nu\ll W$, one obtains $\Gamma_{\rm mf}\propto
\nu$, while for $\nu\gg W$, $\Gamma_{\rm mf}\propto\nu(W/\nu)^{d/\mu}$
\cite{uemura-comm}.

We restricted our treatment here to point clusters with unique moment
$m$ and neglected interactions between the moments. As long as the
cluster sizes are much smaller than the mean distance $n^{-1/d}$,
crossover effects to a Debye like relaxation behavior typical for
Gaussian processes should be of minor importance. A broad distribution
of cluster sizes, however, may require a refined analysis in the
dynamic regime (in the static regime the results remain unchanged
except that $m$ in the width $W$ has to be replaced by its average
value). To capture the dominant contributions to the local field and
to take into account the variation in the jump frequencies (associated
with changes in the anisotropy energy), it can be necessary to define
the subensembles with respect to both the distance of the clusters
nearest to the spin probe and the size of the clusters. Effects due to
dispersion in the jump frequencies have been observed, for example, by
$\mu$SR in colossal magnetoresistive manganites
\cite{Heffner/etal:2001}. Nevertheless, the basic scaling arguments
presented in this work would still be applicable and an extension to
systems of clusters with differing moments should be straightforward.

Interactions between the cluster moments at high temperatures $T$ can
be accounted for by a temperature dependent width $W\!=\!W(T)$ in
(\ref{eq:hdistribution}) (for an approximate calculation in
$\mu\!=\!d\!=\!3$, see \cite{Held/Klein:1975}). At low temperatures
$T$ by contrast, the cluster dynamics cannot be described any longer
by a Poisson process with rate $\nu$ (for dipolar systems in
$d\!=\!2,3$ this occurs for $T\!\lesssim\!0.5\,m^2n^{d/3}$, see
\cite{Pendzig/Dieterich:1996,Rinn/etal:1998}). In this low-temperature
regime the problem becomes more difficult and the relaxation laws
(\ref{eq:sz(t)-slow},\ref{eq:sz(t)-fast}) may no longer hold true.  A
non-Poissonian cluster dynamics has recently been encountered in
a spin glass also \cite{Keren/etal:1996}.

Having mentioned these limits of our findings, we hope that our work
will stimulate further research on the challenging problem of spin
precession in disordered systems. Our scaling methods should give
deeper insight into the spin relaxation in disordered systems and may
be extended to describe $\mu$SR (or $\beta$NMR) in other complex
systems, as e.g.\ spin glasses, structural glasses, amorphous magnets
or disordered superconductors.

We should like to thank W.~Dieterich and Ch.~Niedermayer for
discussions and gratefully acknowledge financial support by the
Sonderforschungsbereich 513 and the Heisenberg program (P.M.)  of the
Deutsche Forschungsgemeinschaft.

\end{multicols}


\begin{references}
\bibitem{mu-d-comm} For $\mu<d/2$, $\langle H^2\rangle$ is finite but
  vanishes in the thermodynamic limit. This yields a Debye like
  relaxation typical for Gaussian processes but the characteristic
  relaxation times would depend on the system size.
\bibitem{c-comm} Explicit known constants referred to in the text are
$C_W^{d/\mu}\!=\!\mu^{-1} S_d \int_0^\infty\!\!du\,
u^{-1-d/\mu}(1\!-\!\sin u/u)$,
$C_\psi\!=\!2\pi^{-1}\Gamma(2\!+\!d/\mu)\sin(\pi d/2\mu)$,
$C_H\!=\!S_d/[(2\mu\!-\!d)C_W^{d/\mu}]$, where
$S_d\!=\!2\pi^{d/2}/\Gamma(d/2)$ and $V_d\!=\!S_d/d$ are the surface and
volume of the $d$-dimensional unit sphere, respectively.
\end{references}
\end{document}